\documentclass[twocolumn,showpacs,preprintnumbers,amsmath,amssymb,prl,superscriptaddress,longbibliography]{revtex4-1}
\usepackage{bbm}
\usepackage{mathrsfs}
\usepackage{graphicx}
\usepackage{dcolumn}
\usepackage{bm}
\usepackage{amsmath}
\usepackage{amsfonts}
\usepackage{color}
\usepackage[colorlinks=true,linkcolor=magenta,urlcolor=magenta,citecolor=cyan,anchorcolor=blue]{hyperref}
\usepackage{floatrow}

\begin{document}
	
\title{Discovering two-dimensional magnetic topological insulators by machine learning}
\author{Haosheng Xu}
\affiliation{State Key Laboratory of Surface Physics and Department of Physics, Fudan University, Shanghai 200433, China}
\author{Yadong Jiang}
\affiliation{State Key Laboratory of Surface Physics and Department of Physics, Fudan University, Shanghai 200433, China}
\author{Huan Wang}
\affiliation{State Key Laboratory of Surface Physics and Department of Physics, Fudan University, Shanghai 200433, China}
\author{Jing Wang}
\thanks{Corresponding author: wjingphys@fudan.edu.cn}
\affiliation{State Key Laboratory of Surface Physics and Department of Physics, Fudan University, Shanghai 200433, China}
\affiliation{Institute for Nanoelectronic Devices and Quantum Computing, Zhangjiang Fudan International Innovation Center, Fudan University, Shanghai 200433, China}
\affiliation{Hefei National Laboratory, Hefei 230088, China}
	
\begin{abstract}
Topological materials with unconventional electronic properties have been investigated intensively for both fundamental and practical interests. Thousands of topological materials have been identified by symmetry-based analysis and \emph{ab initio} calculations. However, the predicted magnetic topological insulators with genuine full band gaps are rare. Here we employ this database and supervisedly train neural networks to develop a heuristic chemical rule for electronic topology diagnosis. The learned rule is interpretable and diagnoses with a high accuracy whether a material is topological using only its chemical formula and Hubbard $U$ parameter. We next evaluate the model performance in several different regimes of materials. Finally, we integrate machine-learned rule with \emph{ab initio} calculations to high-throughput screen for magnetic topological insulators in 2D material database. We discover 6 new classes (15 materials) of Chern insulators, among which 4 classes (7 materials) have full band gaps and may motivate for experimental observation. We anticipate the machine-learned rule here can be used as a guiding principle for inverse design and discovery of new topological materials.
\end{abstract}
	
\date{\today}
	
\maketitle
	
Topological materials are exotic states of matter characterized by topologically nontrivial electronic band structure~\cite{hasan2010,qi2011,kane2005b,bernevig2006c,koenig2007,fu2007,chen2009,fu2011,ando2015,benalcazar2017,liu2014a,xu2015a,lv2015,yang2015,armitage2018,bradlyn2016}. Ever since the birth of the field, widespread efforts from first-principles calculations in synergy with topological band theory have been devoted to identify and catalogue candidate topological materials~\cite{bansil2016,xiao2021}. The recent theoretical frameworks known as topological quantum chemistry~\cite{bradlyn2017,slager2017,elcoro2021} and symmetry indicators~\cite{po2017,watanabe2018,po2020} enable efficient diagnosis of topological materials using only symmetry data of the wavefunction~\cite{fu2007a,song2018}. These symmetry-based methods facilitate fruitful computational searches for topological materials~\cite{zhangt2019,vergniory2019,tang2019a,tang2019b,xu2020,vergniory2022}. However, certain forms of band topology and low-symmetry systems are invisible to symmetry indicators~\cite{po2017}. For example, Chern insulators and time-reversal invariant $Z_2$ topological insulators (TI) without any point group symmetry cannot be diagnosed by symmetry indicators, and their topological nature can be determined only by evaluating the wavefunction-based topological invariant directly, which requires significant computational cost. Moreover, the complicated magnetic structure of materials hinders diagnosis by using magnetic topological quantum chemistry~\cite{xu2020}. Thus, for practical reason, it is highly desirable to develop broadly applicable rules to determine whether a given electronic material is topological.

Recently, machine learning (ML) has become a novel efficient tool for predicting topological materials~\cite{claussen2020,cao2020,liu2021,schleder2021,andrejevic2022,ma2023} and topological invariants~\cite{zhangyi2017,zhangpf2018,scheurer2020}. Among these applications, a heuristic chemical rule for electronic topology diagnosis has been proposed, which does not depend on the crystal symmetry~\cite{ma2023}. Motivated by understanding of chemical bonding from electronegativity as its tendency to attract electrons, they termed a ML numerical value for each element as \emph{topogivity}, which loosely captures its tendency to form topological materials. The heuristic rule for electronic topology of a given material is determined by the sign of weighted average of its elements’ topogivities. New non-symmetry-diagnosable topological materials have been predicted by the heuristic rule and density functional theory (DFT) validation. In spite of these suscesses, their work only involved non-magnetic materials and did not include many transition metal elements which constitute magnetic materials. From the perspective of first-principles calculation, the topology of magnetic materials may depend on Hubbard $U$ parameters~\cite{xu2020}. Thus the dependence of topology on $U$ value cannot be captured by their chemical rule~\cite{ma2023,felser2021,schoop2018,cava2019}. Meanwhile, the number of confirmed magnetic topological materials is less than ten~\cite{bernevig2022}. This motivate us to develop ML chemical rule for efficient electronic topology diagnosis and searches for magnetic materials. 

\begin{figure}[t]
\begin{center}
\includegraphics[width=3.4in, clip=true]{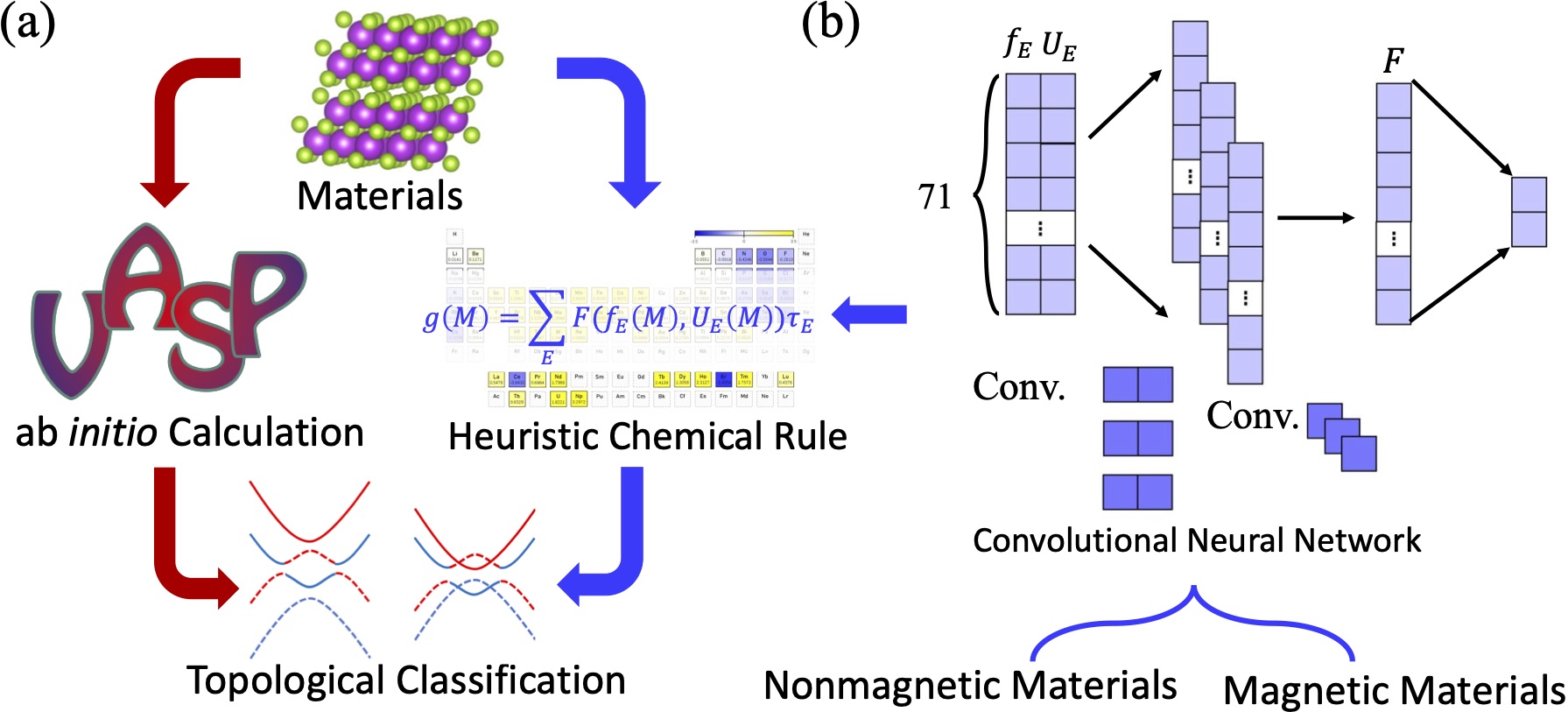}
\end{center}
\caption{Heuristic chemical rule diagnosis and DFT discovery of topological materials. (a) The topogivity-based heuristic diagnosis of a given material is evaluated by weighting the material's elements' topogivities $\tau_E$ with $\mathcal{F}(f_E,U_E)$, where $\mathcal{F}(f_E,U_E)$ is determined by element's fraction $f_E$ in the chemical formula and Hubbard parameter $U_E$. The high-throughput search for topological materials is performed by rapid heuristic rule screening through the material database to get candidate topological materials, and then followed by DFT calculations~\cite{kresse1996}. (b) Schematic of the ML workflow and structure of CNN, where the heuristic chemical rule is learned with both non-magnetic and magnetic materials as input.}
\label{fig1}
\end{figure}

\begin{figure}[t]
\begin{center}
\includegraphics[width=3.4in, clip=true]{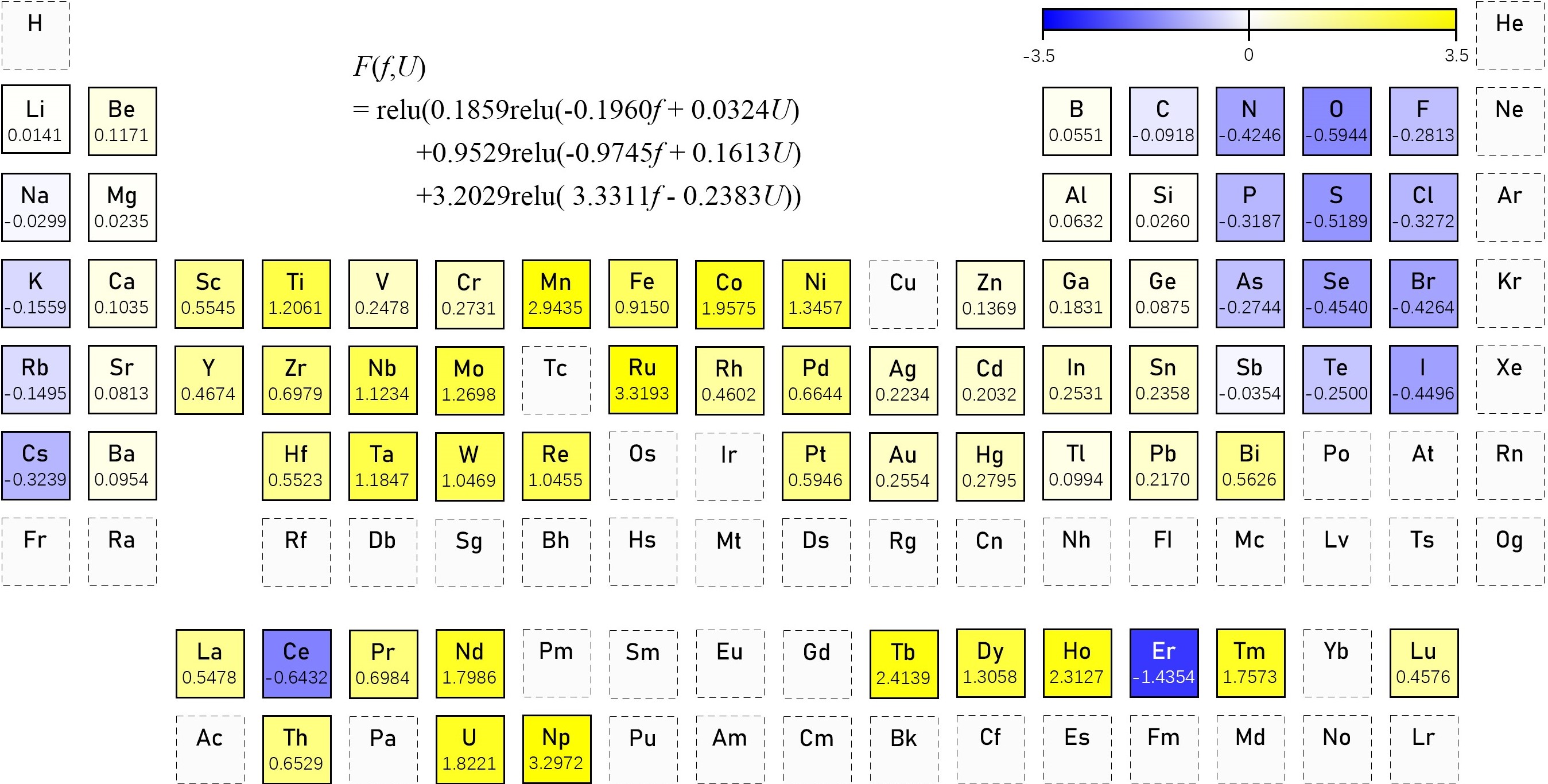}
\end{center}
\caption{Periodic table of ML topogivities $\tau_{E}$ and combinatorial weight $\mathcal{F}(f,U)$. $\tau_{E}$ are shown by color-coding and in values. Elements that  appear less than 15 times in the labeled dataset are shown in gray with dashed box.}
\label{fig2}
\end{figure}

\begin{figure}[t]
\begin{center}
\includegraphics[width=3.4in, clip=true]{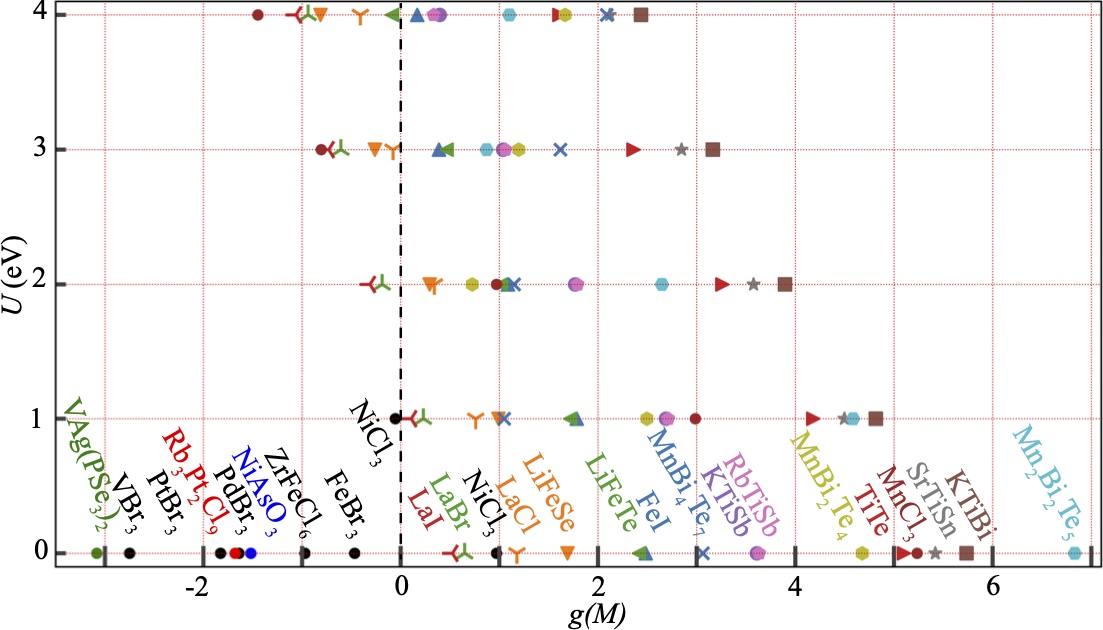}
\end{center}
\caption{The computed $g(M)$ vs $U$ for Chern insulators, which were predicted by first-principles calculations under certain Hubbard $U$ parameters. Materials with $g(M)<0$ at $U=0$ is shown only, for they are even more negative at finite $U$.}
\label{fig3}
\end{figure}

Here, as illustrated in Fig.~\ref{fig1}, we use the convolutional neural network (CNN) to search for chemical rules of topological electronic structure by including Hubbard $U$ value. We obtain training parameters $\tau_E$ (referred as topogivity) for each element in the periodic table, and find the heuristic rule of a given material is diagnosed with high accuracy (average 83.9\%) as topologically nontrivial (trivial) if the weighted average of its elements’ topogivities is positive (negative). Here the element weight of a given material is determined by both the element's fraction and Hubbard $U$ parameter (Fig.~\ref{fig2}). The convolution layers correctly capture the influence of $U$ value on the topological properties of magnetic materials displayed in the training set. We first test our heuristic rule to predict non-symmetry-diagnosable and non-magnetic topological materials as in Ref.~\cite{ma2023} and get very similar accurate results. Then we proceed to perform model evaluation in Chern insulators~\cite{zhang2019,li2019,otrokov2019a,liyue2020,sunhy2019,liy2020,sun2020,xuan2022,jiang2023,dolui2015,liuzhao2018,he2017,sun2018,you2019,sunj2020,lizy2021,lizy2022,choudhary2020}, and find our heuristic rule for diagnosis still has a high balanced accuracy $\sim$82.8\% (Fig.~\ref{fig3}). Finally, we integrate ML rule with DFT calculations to search for magnetic TI in 2DMatPedia~\cite{zhou2019}, and find T-phase RuO$_2$, OsO$_2$, GdBr and Tb$X$ family are new Chern insulators with full band gaps.

{\color{cyan}\emph{Training and testing dataset.}} 
We employ a supervised learning to obtain heuristic chemical rule for topological materials diagnosis. Here the training dataset consists of nonmagnetic and magnetic, stoichiometric, three-dimensional materials, where we label TIs, topological crystalline insulators and topological semimetals (TSM) as topological materials, and refer all other materials as trivial materials. The nonmagnetic dataset utilizes a subset of the database developed in Ref.~\cite{tang2019a}, where only the space groups with nontrivial symmetry indicator groups are taken~\cite{supple}. We add the data of magnetic materials identified in Ref.~\cite{xu2020}, where the same material with transition metal element of different $U$ values may belong to different topological classifications. For instance, Mn$_{5}$Si$_{3}$ is a TI with $U=0$ for Mn, a TSM with $U=1$~eV, and trivial with $U=2, 3, 4$~eV. Here we consider a given magnetic material with different Hubbard $U$ parameters as different inputs, which further expands our magnetic training data (see Supplementary Material for methodology of constructing the training dataset). Then our labeled dataset comprises 9284 materials, of which $51.8\%$ are marked as topological (69.5\% are TSM) and the remaining $48.2\%$ are marked as trivial. However, it is worth noting that certain topology may not be correctly identified by symmetry-based methods, thus the training dataset should be viewed as a set with noisy labels. The evaluation of our  model is performed in several different settings which are not contained in training dataset. 

{\color{cyan}\emph{Heuristic chemical rule and CNN.}} 
Our ML heuristic chemical rule takes the form
\begin{equation}\label{rule}
g(M)=\sum_{E}\mathcal{F}(f_{E}(M),U_{E}(M))\tau_{E},
\end{equation}
where the summation runs over all elements in material $M$, $\tau_{E}$ is a learned parameter for element $E$, and $\mathcal{F}$ is learned by the convolution layers (Fig.~\ref{fig2}), which is a function of $f_{E}(M)$ and $U_{E}(M)$. Here $f_{E}(M)$ is the element fraction for element $E$ in material $M$ (e.g., for a chemical formula $X_aY_bZ_c$, $f_{X}(M)=\frac{a}{a+b+c}$, $f_{Y}(M)=\frac{b}{a+b+c}$, $f_{Z}(M)=\frac{c}{a+b+c}$), and $U_{E}(M)$ is Hubbard $U$ value for element $E$ (if the element is non-magnetic, we set it to zero). The sign of $g(M)$ decides the classification: classify as topological (trivial) if $g$ is positive (negative). A larger value of $g(M)$ roughly corresponds to a more confident classification decision. Thus a diagnosis is obtained only by material’s chemical formula and Hubbard $U$ parameters.

The structure of CNN is shown in Fig.~\ref{fig1}(b). A material is described by a $71\times2$ matrix with each row representing an element of periodic table. The first and second columns represent the element's fraction and $U$ value of the material, respectively. The convolutional network has two convolutional layers with $3$ kernels of size $1\times2$ and $1$ kernel of size $1\times1$, followed by a binary classification neural network. The total number of trainable parameters is $151$. All the hidden layers have rectified linear units $\text{relu}(x)=\text{max}\{0,x\}$ as activation functions. The output layer has softmax activation function given by the shape of $(\frac{e^A}{e^A+e^B}, \frac{e^B}{e^A+e^B})^T$, where $A=\sum_{E}\mathcal{F}_{E}a_{E}$ and $B=\sum_{E}\mathcal{F}_{E}b_{E}$. The model is trained by marking trivial material as $(1, 0)^T$ and topological material as $(0, 1)^T$. The network produces two sets of learning parameters $a_{E}$ and $b_{E}$ for each element $E$, we find the material is judged to be topological when $\frac{e^B}{e^A+e^B} > \frac{e^A}{e^A+e^B}$, which is equivalent to $B-A=\sum_{E}\mathcal{F}_{E}(b_{E}-a_{E})\equiv\sum_{E}\mathcal{F}_{E}\tau_{E}>0$.

It is interesting to compare our learned chemical rule to that learned by support vector machine in Ref.~\cite{ma2023} which applies to non-magnetic materials only. For non-magnetic materials with $U=0$, $\mathcal{F}(f_E(M),0)\propto f_{E}(M)$, then Eq.~(\ref{rule}) reduces to $g(M)\propto\sum_{E}f_{E}(M)\tau_{E}$, which is exactly the same heuristic rule learned in Ref.~\cite{ma2023}. While for magnetic elements with finite $U$, $\mathcal{F}$ is no longer a simple function of $f_E(M)$, and $\mathcal{F}(f_E(M),U_E(M))<\mathcal{F}(f_E(M),0)$, namely Hubbard $U$ value reduces the weighting of corresponding magnetic element. Thus the topogivity $\tau_E$ for each element $E$, loosely captures the tendency of an element to form topological materials.

We first evaluate our model performance within the labeled dataset before making predictions in different settings. We did eight-fold cross-validation and averaged the results over multiple test sets, and found an average of 83.9\% accuracy. Moreover, we find empirically the fraction of accurately classified materials increases as the value of $g(M)$ increases. The accuracy reaches 93\% when $|g(M)|\approx3.5$, after which the accuracy does not increase significantly~\cite{supple}. Specifically, we observe on average that $94.8\%$ of materials with $g(M)\geq4.0$ are correctly classified to be topologically nontrivial. Having completed the cross validation, we use the entire labeled dataset to fit the final model, which is what we will use for making predictions in different regimes. We observed the balanced accuracy for the magnetic materials only in the training dataset is $82.8\%$. Additionally, we found that the balanced accuracy of the model is better for materials with two or three distinct elements than for materials with one or four distinct elements.
		
Our model’s learned topogivities and weightings $\mathcal{F}$ are shown in Fig.~\ref{fig2}, where the elements that  appear less than 15 times in the labeled dataset are shown in gray (see the full table of elements' topogivities and appearing times in Supplementary Material). This table of topogivities enables a fast heuristic diagnosis of any stoichiometric material whose elements are featured in the periodic table. For example, magnetic TI MnBi$_2$Te$_4$~\cite{zhang2019,gong2019,otrokov2019} does not appear in the labeled dataset, Weyl semimetal TaAs~\cite{xu2015a,lv2015} is non-symmetry-diagnosable, but both of them are successfully diagnosed as topological by our learned rule: $g(\text{MnBi}_2\text{Te}_4)=1.195$ ($U=3$~eV for Mn) and $g(\text{TaAs})=4.86$.

The specific learned value of element topogivities are in general affected by the dataset and modeling limitations. However, several chemical heuristics can be extracted from the table of topogivities qualitatively. First, similar to Ref.~\cite{ma2023}, two clusters of elements located in the top right and bottom left parts of the periodic table have negative topogivities, which is consistent with intuition, since these two clusters tend to form ionic crystals and often have trivial band  gaps. Second, considering groups 13 to 16, the topogivity decreases as one move from left to right across a period and increases as one move down a group, which is opposite to the electronegativity trend in the periodic  table. This is also consistent with intuition that heavier elements have larger spin-orbit coupling and weaker electronegativities form covalent crystals with smaller band gap, both of them often play important roles in topological materials. Finally, we observe all transition metals have positive topogivities. As Hubbard $U$ increases, the weighting of transition metal elements decreases in a material, which leads to decreasing of $g(M)$. This is consistent with the intuition that large $U$ value often lead to Mott insulator. Overall, these chemical insights suggest the topogivity-based picture and heuristic rule can provide a useful way to study topological materials. 

{\color{cyan}\emph{Evaluating the rule in different settings.}} We then evaluate our model in different regimes of materials compared to the labeled dataset. First, we apply the learned rule and compute $g(M)$ for materials in the discovery space, which contains $1431$ non-symmetry-diagnosable and non-magnetic materials~\cite{ma2023,supple}. We set a threshold of $1.6$ for $g(M)$ which corresponds to a high-confidence topological nontrivial classification, and leaves 79 materials. We further eliminate 7 materials which contain 4f or 5f electron with 72 materials left for DFT validation. We perform DFT within generalized-gradient approximation, and include spin-orbital coupling~\cite{supple}. Of the 72 materials, we find 62 topological materials, corresponding to a success rate of $86.1\%$. All of the 62 topological materials are TSM, where 55 materials are consistent with the finding in Ref.~\cite{ma2023}. Among the remaining 7 topological materials that we identified here, 3 have been predicted previously in the literature and the rest 4 represent truly new materials discovery~\cite{supple}.

Second,  we use the final model to compute $g(M)$ for trivial magnetic materials identified in Ref.~\cite{xu2020}. There are 200 such materials (after the remove of 41 materials containing elements without topogivities) which are trivial at any $U$ value. The test dataset is generated by combining these materials with different $U$ values. We find that the model classifies $77.6\%$ of materials in this set as trivial. It is interesting to compare this $77.6\%$ number to the specificity, which is the fraction of samples classified as trivial among all the samples that have a label of trivial. We observe some deterioration in model performance, where the test specificity is $85.6\pm1.6\%$ in the cross validation process.

At last, we evaluate our model performance to Chern insulators predicted by DFT calculations~\cite{zhang2019,li2019,otrokov2019a,liyue2020,sunhy2019,liy2020,sun2020,xuan2022,jiang2023,dolui2015,liuzhao2018,he2017,sun2018,you2019,sunj2020,lizy2021,lizy2022,choudhary2020}. There are quite a few classes of 2D Chern insulator materials with full band gaps. For example, thin film of intrinsic magnetic TI family Mn$_m$Bi$_{2n}$Te$_{m+3n}$~\cite{zhang2019,li2019,otrokov2019a,liyue2020,sunhy2019}, FeI and TiTe~\cite{liy2020,sun2020,xuan2022,jiang2023}, LaX~\cite{dolui2015,liuzhao2018}, transition metal trihalides MX$_3$~\cite{he2017,sun2018,you2019,sunj2020}, etc. The test dataset is generated by these materials with different $U$ values~\cite{supple}, since the Chern insulators would be trivial under certain Hubbard $U$ parameters from DFT. This individual dataset is heavily imbalanced in terms of the ratio of topological labels to trivial labels. The computed $g(M)$ vs $U$ for these 2D magnetic materials is shown in Fig.~\ref{fig3}. $g(M)$ is a decreasing function of $U$, but not always monotonic. We find the balanced accuracy of our final model is $82.8\%$. We stress that the validation dataset and the labeled dataset correspond to different regimes of materials, and so it is quite interesting that a model that was fit on the labeled dataset of 3D materials still works in the validation dataset of 2D materials. Additionally, we observe the misclassification in MX$_3$ and Rb$_{3}$Pt$_{2}$Cl$_{9}$, because the halogen with negative topogivities have a dominant fraction in the chemical formula, while their orbitals are far away from Fermi energy and do not contributes to topological bands.

{\color{cyan}\emph{High throughput screening of 2D material and first-principles calculation.}}
Finally, we employ the topogivity-based chemical rule to identify 2D magnetic TI with genuine full band gaps, which are \emph{extremely rare} in the literatures compared to 3D TSM and TI. We compute $g(M)$ for each of 6351 2D materials from 2DMatPedia~\cite{zhou2019}, and found $28\%$ has positive $g(M)$. We focus our attention to materials that have a $g(M)\geq6$ value at $U=0$ that corresponds to a topologically nontrivial classification with high-confidence: that leaves $234$ materials listed in Supplementary Materials. We then eliminate 137 materials without magnetic moment with 97 left for DFT validation. We find 16 magnetic topological materials, among which 11 are TSM~\cite{supple} and the rest 5 (Tb$X$, RuO$_2$) are new classes of Chern insulators listed in Table~\ref{chern_insulator}. To our knowledge these have not been previously predicted. We also find topological band inversion with SOC induced gap in TaCoTe$_2$, which has been predicted and experimentally observed~\cite{lisi2019,mazzola2023}. Furthermore, we expand the search list and set a threshold of $6>g(M)\geq0$, and find 18 Chern insulators, where 8 have been predicted previously~\cite{supple} shown in Fig.~\ref{fig3} and only MnBi$_{2}$Te$_{4}$ has already been experimentally observed. The rest 10 are new material discovery listed in Table~\ref{chern_insulator}, where OsO$_2$ and GdBr has full band gap, Sc$X$ and Y$X$ are metallic but have nontrivial Wilson loop~\cite{supple}.

\begin{table}[t]
\caption{Newly discovered Chern insulators by the heuristic chemical rule and first-principles calculations. {$X$=F, Cl, Br, I. Here $g(M)$ is computed with $U=2, 4, 1, 0$~eV for Ru, Tb, Os, and Gd (Sc, Y), respectively. The first 4 classes have full band gaps, while the remaining 2 classes have Fermi pockets. The topogivities of Os and Gd are in Supplementary Materials.}}
\begin{center}\label{chern_insulator}
\renewcommand{\arraystretch}{1.3}
\begin{tabular*}{3.4in}
{@{\extracolsep{\fill}}cccc}
\hline
\hline
Materials & 2DMat id (2dm-) & $g(M)$ & $\mathcal{C}$  \\
\hline
Tb$X$ & 959, 3600, 3487, 225 & 4.4, 4.1, 3.6, 3.5 & $-1$ \\
RuO$_2$ & 6443 & 2.5 & $2$ \\
OsO$_2$ & 3912 & 1.7 & $2$ \\
GdBr & 5865 & 0.1 & $-1$ \\
\hline
Sc$X$ & 1139, 3544, 1219, 1254 & 1.5, 1.2, 0.7, 0.6 & $-1$ \\
Y$X$ & 984, 4695, 870, 3198 & 1.0, 0.7, 0.2, 0.1 & $-1$ \\
\hline
\hline
\end{tabular*}
\end{center}
\end{table}

\begin{figure}[b]
\begin{center}
\includegraphics[width=3.4in]{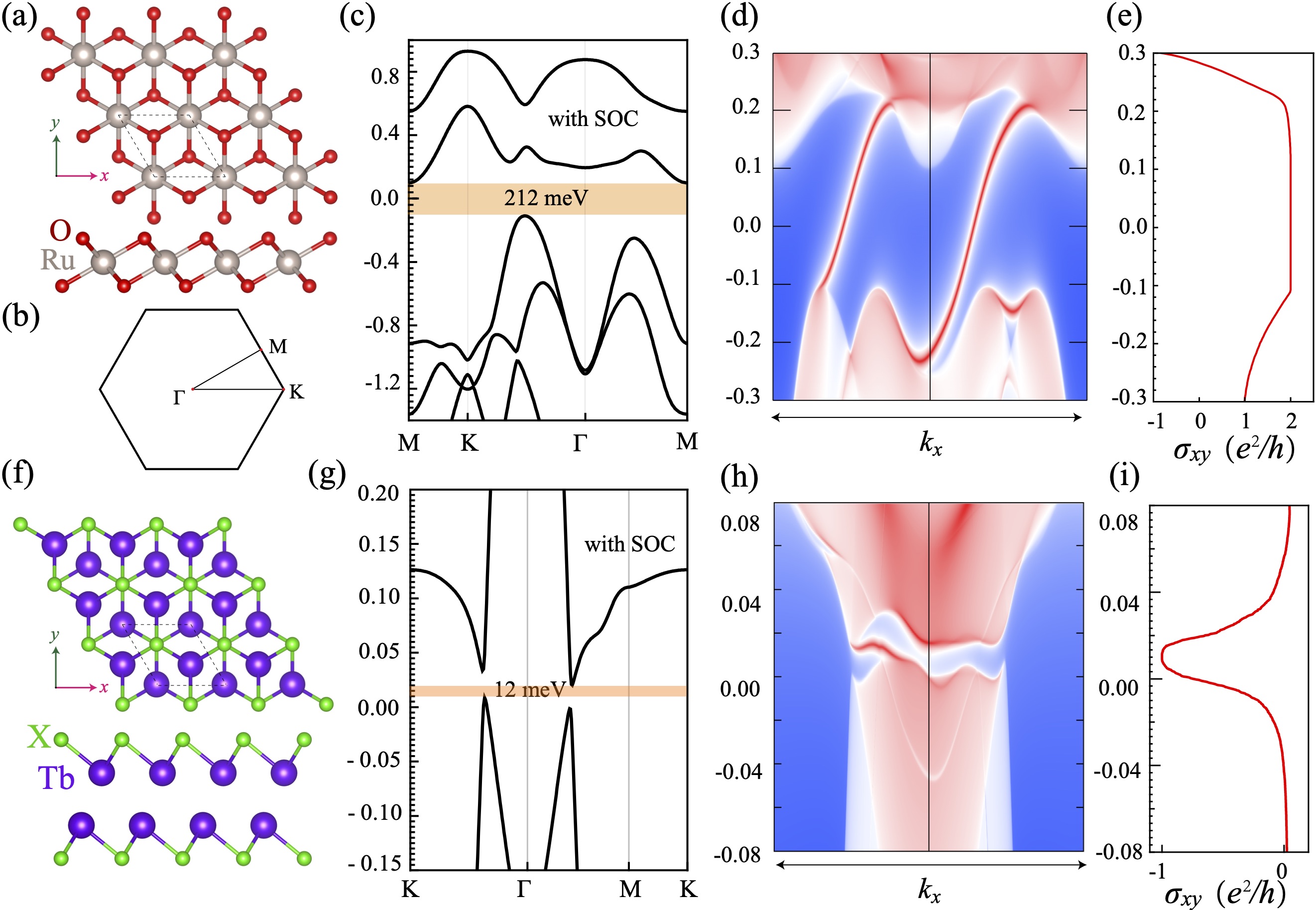}
\end{center}
\caption{Electronic structure and topological properties of monolayer RuO$_2$ and TbBr by DFT+$U$ ($U$=2, 4 eV for Ru-$d$, Tb-$f$ orbital, respectively.). (a)-(e) RuO$_2$, (f)-(i) TbBr, The top and side views of atomic structure; the band structure with SOC; topological edge states calculated along $x$ axis; anomalous Hall conductance $\sigma_{xy}$ as a function of Fermi energy. (b) Brillouin zone. The shaded regions in (e) and (k) denote the topological gap.}
\label{fig4}
\end{figure}
		
We highlight two particularly interesting newly discovered Chern insulators in Fig.~\ref{fig4}. Both of them have topological nontrivial full band gaps, making it
promising for potential experimental investigation. T-phase RuO$_2$ has a hexagonal lattice with space group $P$-$3m1$ (No.~164). Its monolayer was predicted to be unstable and Peierls distorted into T$^\prime$-phase~\cite{ersan2018}. However, recent experiment has observed stable T-phase RuO$_2$ when fabricating H-phase RuO$_2$ nanosheets~\cite{ko2018}. The topology is from spin up band of $d_{z^2}$ orbitals and spin down band of $d_{xz,yz}$ orbitals of Ru at $\Gamma$ point, which leads to $\mathcal{C}=2$. Tb$X$ also form a hexagonal lattice with space group $P$-$3m1$ (No.~164), the topology is from spin down band of $d_{z^2}$ orbitals and spin up band of $d_{xy,x^2-y^2}$ orbitals of Tb at $\Gamma$ point, and leads to $\mathcal{C}=-1$. Detailed analysis on topology of these two systems are in Supplementary Materials.

{\color{cyan}\emph{Discussion.}}
The topogivity-based approximate picture of Eq.~(\ref{rule}) provides a simple but coarse-grained approach for topological materials diagnosis with high accuracy, using only its chemical composition and Hubbard $U$ value, without costly DFT calculations. Our final model cannot guarantee that a real material has topological features, which must be validated by DFT calculation. Still, it provides a fast and efficient tool to classify topological nature of a given material. The magnetic materials dataset helps us to get topogivity data on vast number of transition metal elements. We observe that materials in general with a large number of $d$- or $f$-shell valence electrons, and compounds containing heavy elements with strong spin-orbit coupling, have a greater tendency to be topological nontrivial.

Future research should try to take into account the relative electronegativity of the elements in the compounds. Many misclassified materials (with three or more distinct elements) have element taking a large fraction, but only acting as anion and not contributing to topological band around Fermi energy. Also it is necessity to verify the heuristic rule with more advanced graph neural network by taking materials' crystal symmetry into account. Then it is important to fully understand why the heuristic chemical rule here works so well, which may further elucidate the fundamental question of why some materials are topological while others are not. Furthermore, it is interesting to perform more comprehensive searches and inverse design for new magnetic topological materials using our learned model.

\begin{acknowledgments}
{\color{cyan}\emph{Acknowledgment.}} This work is supported by the National Key Research Program of China under Grant No.~2019YFA0308404, the Natural Science Foundation of China through Grant No.~12174066, the Innovation Program for Quantum Science and Technology through Grant No.~2021ZD0302600, Science and Technology Commission of Shanghai Municipality under Grant No.~20JC1415900, Shanghai Municipal Science and Technology Major Project under Grant No.~2019SHZDZX01.
\end{acknowledgments}

\end{document}